\documentclass[final]{svjour3}
\usepackage{graphicx}
\usepackage{rotating}
\usepackage{amssymb}
\usepackage[numbers]{natbib}
\usepackage{wrapfig}
\usepackage{float}
\usepackage{url}

\usepackage{graphicx}
\usepackage[numbers]{natbib}
\makeatletter
\journalname{Journal of Low Temperature Physics}

\bibpunct{}{}{,}{s}{}{,}

\begin{document}

\title{Design of the on-board data compression for the bolometer data of LiteBIRD}
\titlerunning{Design of the on-board data compression for LiteBIRD}

\author{
Mayu Tominaga \and
Masahiro Tsujimoto \and
Graeme Smecher \and
Hirokazu Ishino
on behalf of LiteBIRD Joint study group
}
\authorrunning{Tominaga et al.}

\institute{
Mayu Tominaga \at
The University of Tokyo, Graduate School of Science, Department of Astronomy, 7-3-1 Hongo, Bunkyo-ku, Tokyo 113-8654 Japan.
\email{tominaga@ac.jaxa.jp} \and
Masahiro Tsujimoto \at
Japan Aerospace Exploration Agency, Institute of Space and Astronautical Science, Department of Space Astronomy and Astrophysics, 
3-1-1 Yoshino-dai, Chuo-ku, Sagamihara, Kanagawa 252-5210, Japan. \and
Graeme Smecher \at
Three-Speed Logic, Inc., 
Victoria, B.C., V8S 3Z5, Canada \and
Hirokazu Ishino \at
Okayama University, Department of Physics, 
3-1-1, Tsushima-Naka, Kita-ku, Okayama-shi, 700-8530 Japan.
}

\date{Received: date / Accepted: date}

\maketitle

\begin{abstract}
LiteBIRD is a space-borne experiment dedicated to detecting large-scale
$B$-mode anisotropies in the linear polarization of the Cosmic Microwave Background
(CMB) predicted by the theory of inflation.  It is planned to be launched in the late
2020s to the second Lagrange point (L2) of the Sun-Earth system. LiteBIRD will
map the sky in 15 frequency bands.  In comparison to \textit{Planck} HFI, the previous
low-temperature bolometer-based satellite for CMB observations, the number of detector
has increased by two orders of magnitude, up to $\sim$5000 detectors in total. The data
rate is 19~Hz from each detector. The bandpass to the ground is limited to 10~Mbps using
the X-band for a few hours per day. These require the data to be compressed by more than
50\%. The exact value depends on how much information entropy is contained in the real
data. We have thus evaluated the compression by simulating the time-ordered data of
polarization sensitive bolometers. The foreground emission, detector noise, cosmic ray
glitches, leakage from the CMB intensity to polarization, etc. are simulated. We
investigated several algorithms and demonstrated that the required compression ratio can
be achieved by some of them. We describe the details of this evaluation and propose
algorithms that can be employed in the on-board digital electronics of
LiteBIRD.

\end{abstract}

\section{Introduction}\label{s1}
LiteBIRD is a satellite dedicated to observing the anisotropy of the linear
polarization of the Cosmic Microwave Background (CMB). It aims to detect the $B$-mode
pattern at a large angular scales ($\ell<200$) to constrain the theory of inflation. The
mission is expected to achieve the sensitivity of the tensor-to-scalar ratio $\delta r
\leq 0.001$ in order to discriminate physically well-motivated models. The satellite is
planned to be launched in the late 2020s by JAXA and is designed by a collaboration of
many institutions worldwide \cite{Hazumi2020}.

LiteBIRD has three telescopes (LFT \cite{Sekimoto2020}, MFT, and HFT \cite{Montier2020})
covering 15 frequency bands over 34--448~GHz. They have in total $\sim$~5000 Transition
Edge Sensor (TES) bolometers cooled at 100~mK. Increasing the number of bolometers is
essential in modern CMB experiments dominated by photon noise. In space-borne
experiments, the downlink rate from the spacecraft to the ground stations limits the
entire data rate. The data compression is thus one of the major functional requirements
for onboard data processing units. The algorithm should be developed for each experiment to
exploit the features of the bolometer data. Significant efforts were made in \textit{Planck} HFI
\cite{Planck2018} and LFI \cite{Maris2000a,Gaztanaga2001,Maris2004} with the trade-off
between the cosmological information and the available telemetry rate. This article presents a study on
the onboard data compression algorithm for LiteBIRD.

\section{Bolometer data}\label{s2}
\begin{figure}[hbtp]
 \centering
 \includegraphics[keepaspectratio,width=\columnwidth]{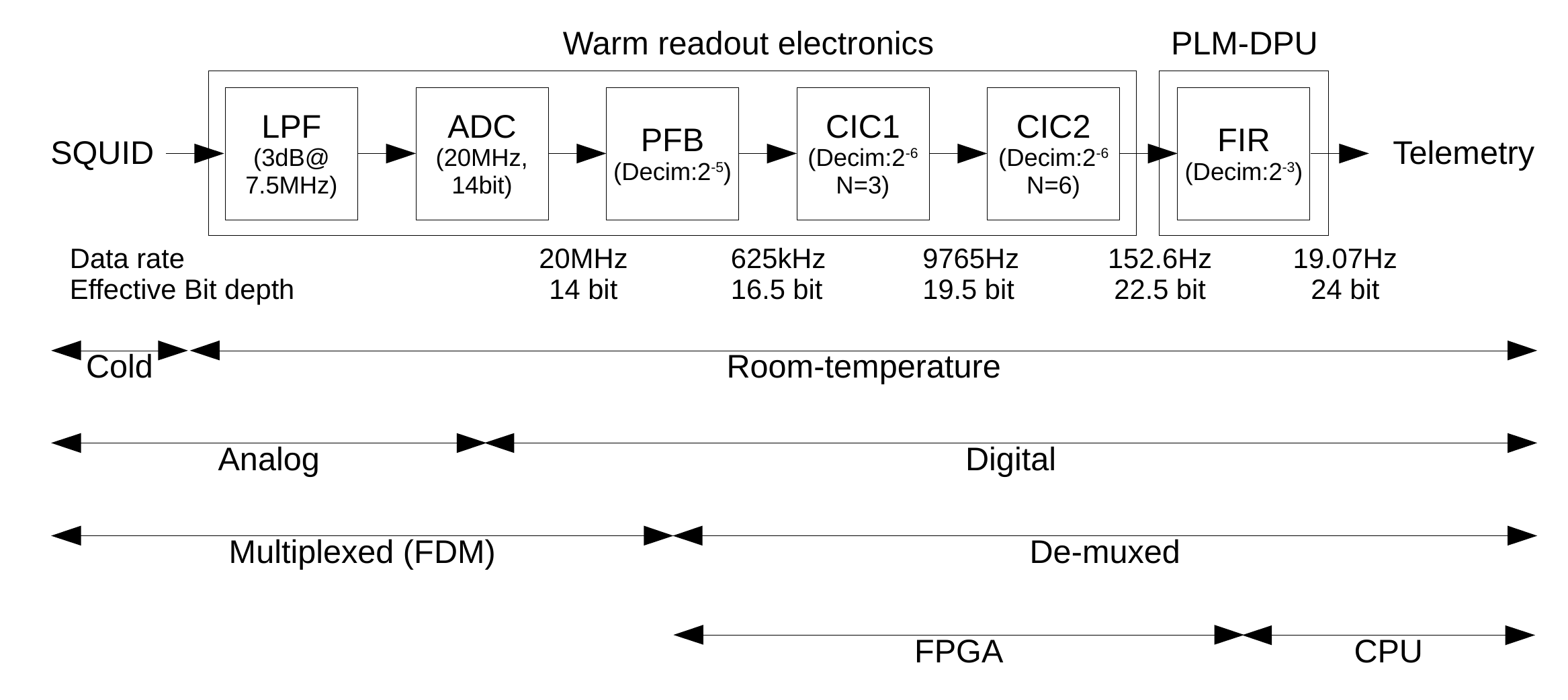}
 \vspace*{-3mm}
 \caption{Onboard digital processing.}
 \label{f1}
\end{figure}

LiteBIRD has $\sim$5000~TES bolometers. The TES detectors are operated at a
bath temperature ($\sim$100~mK) under the constant photon loading by the 2.7~K CMB. The
differential power is cancelled via thermo-electric feedback with a 3.3~ms time constant
\cite{Ghigna2020}. The detectors are read out using SQUIDs by digital frequency-domain
multiplexing \cite{Dobbs2012,Montgomery2020}, using an anti-aliasing filter at
$\sim$7.5~MHz followed by sampling at 20~MHz ($=$ 20~M samples/s). The rate is decimated
by a factor of 2$^{20}$ on board using a series of digital filters consisting of a
poly-phase filter bank (PFB), two cascaded integrated comb (CIC) filters, and finite
impulse response (FIR) filter using two hardware units at room temperature
(Fig. \ref{f1} left). The warm readout electronics (WRE) is mainly responsible for fast
data processing down to 20M/2$^{17}=153$~Hz using an FPGA, while the payload module data
processing unit (PLM-DPU) is for slow and LiteBIRD-specific processing down to
20M/2$^{20}\approx19$~Hz using an FPGA and a CPU. A further downsampling would cause an
unacceptable level of systematics in the modulated signal. The data compression
algorithm is implemented in the PLM-DPU.

The ADC has a 13 bit resolution (bipolar, 14 bit). The effective bit length increases by
10 bits after a 2$^{20}$ decimation. The downlinked data thus have 24~bits of information. The
approximate total rate is 5000~detectors $\times$ 24~bit depth $\times$ 19~Hz rate
$\times$ 24~hours $=$ 25.9~GB day$^{-1}$. The available downlink rate is 10~MHz
(X-band communication) $\times$ a few (3) hours of contact $=$ 13.5~GB day$^{-1}$. We thus
need to achieve a 50\% data compression ratio. We prefer to compress the data losslessly,
so as not to produce uncontrolled systematic terms.

The incoming signal is modulated by the continuously rotating half wave plate (HWP) at
$f_{\mathrm{HWP}}=$46, 39, and 61~rpm $=$0.77, 0.65, and 1.02 Hz (LFT, MHT, and HFT,
respectively). The Stokes $U$ and $Q$ components are modulated at $4f_{\mathrm{HWP}}$,
whereas a part of the $I$ component is modulated at $2f_{\mathrm{HWP}}$ due to leakage from $T$
($I$ for CMB) to $P$olarization by the HWP imperfections (T-to-P leakage). The signal is
also modulated by the scanning of the satellite, which combines a spin of 50 degrees
in 20~min and a precession of 45 degrees in 3.2058~hr. The entire sky is scanned in
half a year as the L2 point rotates around the Sun.

\section{Simulation}\label{s3}
\begin{figure}[hbtp]
  \begin{center}
  \includegraphics[keepaspectratio,width=\columnwidth]{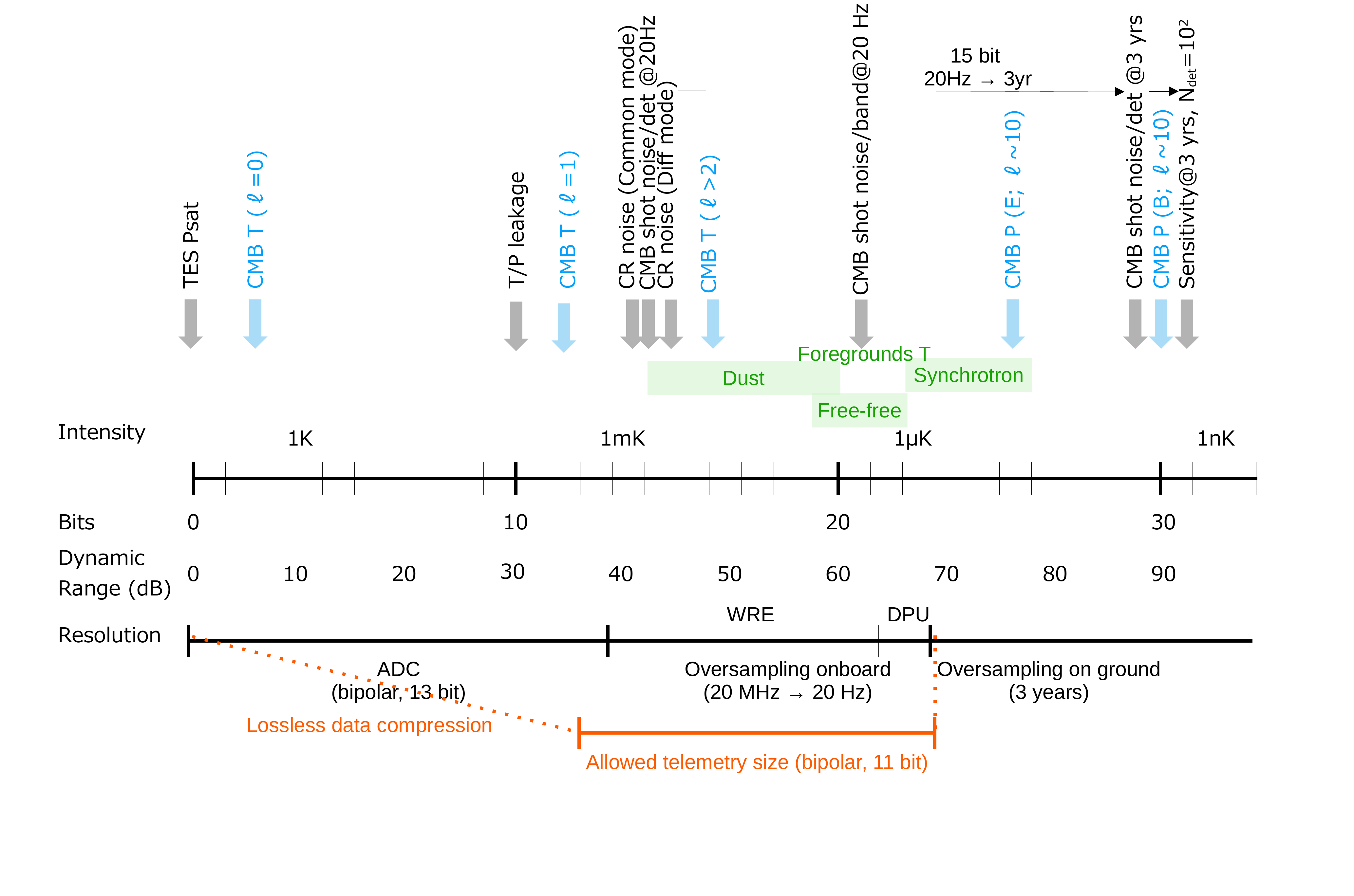}
  \end{center}
  \vspace*{-10mm} 
  \caption{Dynamic range of the experiment. For  cosmic-ray (CR) noise, see Stever et
 al. (2021)\cite{Stever2021}. For CMB shot noise, a noise effective temperature (NET) is assumed to be observed at 100~GHz.}
  \vspace*{-6mm} 
 \label{f2}
\end{figure}

The compression ratio depends on how much information entropy is contained in the
incoming data. This is primarily governed by how we allocate the dynamic range to the
24-bit length. Fig.~\ref{f2} depicts the dynamic range. We set the maximum to
accommodate the saturation power ($\times$2.5 of the optical loading) of TES with an
ample ($\times$5) margin, which corresponds to 1.9--4.8~pW. For 24~bits, the resolution
is 0.22--0.58~aW~bit$^{-1}$ or $\sim$0.8--0.9~$\mu$K$_{\mathrm{CMB}}$ bit$^{-1}$.


We then simulated time ordered data (TOD). The foreground components
\cite{Adam2016b} (synchrotron, dust, free-free, and anomalous microwave emission) and
the CMB (anisotropy) are given in the map domain using \texttt{PySM3} \cite{Thorne2016}
based on \textit{Planck} observations. We mock-observed the sky using the bolometers in the
actual focal plane layout modulated by the satellite scans and the rotating HWP. The CMB
dipole was also simulated. We generated the TOD sampled at approximately 153~Hz using
\texttt{TOAST}\footnote{\url{https://github.com/hpc4cmb/toast}}. They are discretized to
24~bit integers. In the time domain, we
add the detector white noise with a $1/f$ component, cosmic-ray (CR) glitches
\cite{Tominaga2020,Stever2021}, and T-to-P leakage with the assumed leakage ratio of
10$^{-3}$.

\begin{figure}[hbtp]
\vspace*{-3mm}
 \begin{tabular}{cc}
  \begin{minipage}[t]{0.45\hsize}
   \centering
   \includegraphics[keepaspectratio,width=\columnwidth]{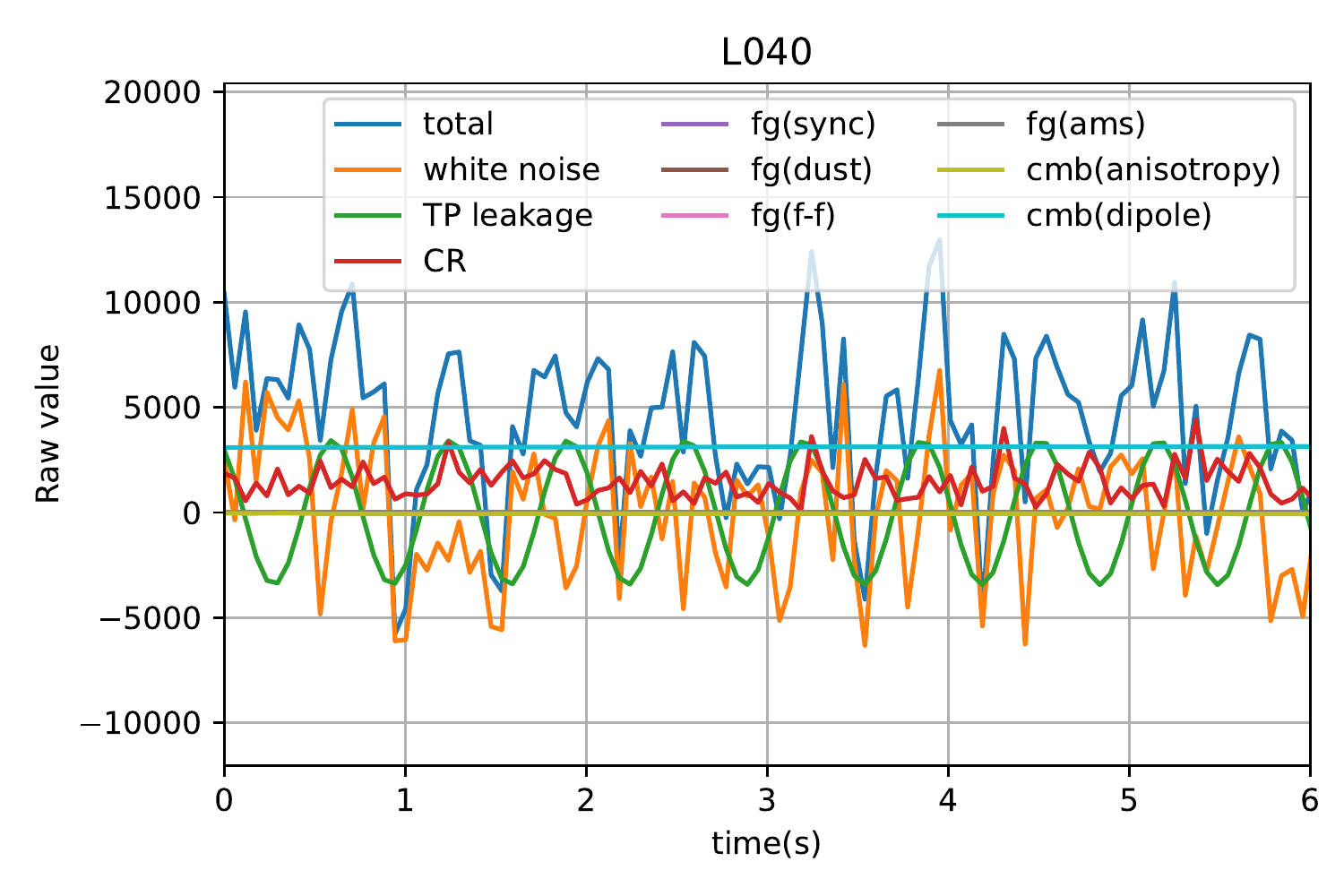}
  \end{minipage}&
  \begin{minipage}[t]{0.45\hsize}
   \centering
   \includegraphics[keepaspectratio,width=\columnwidth]{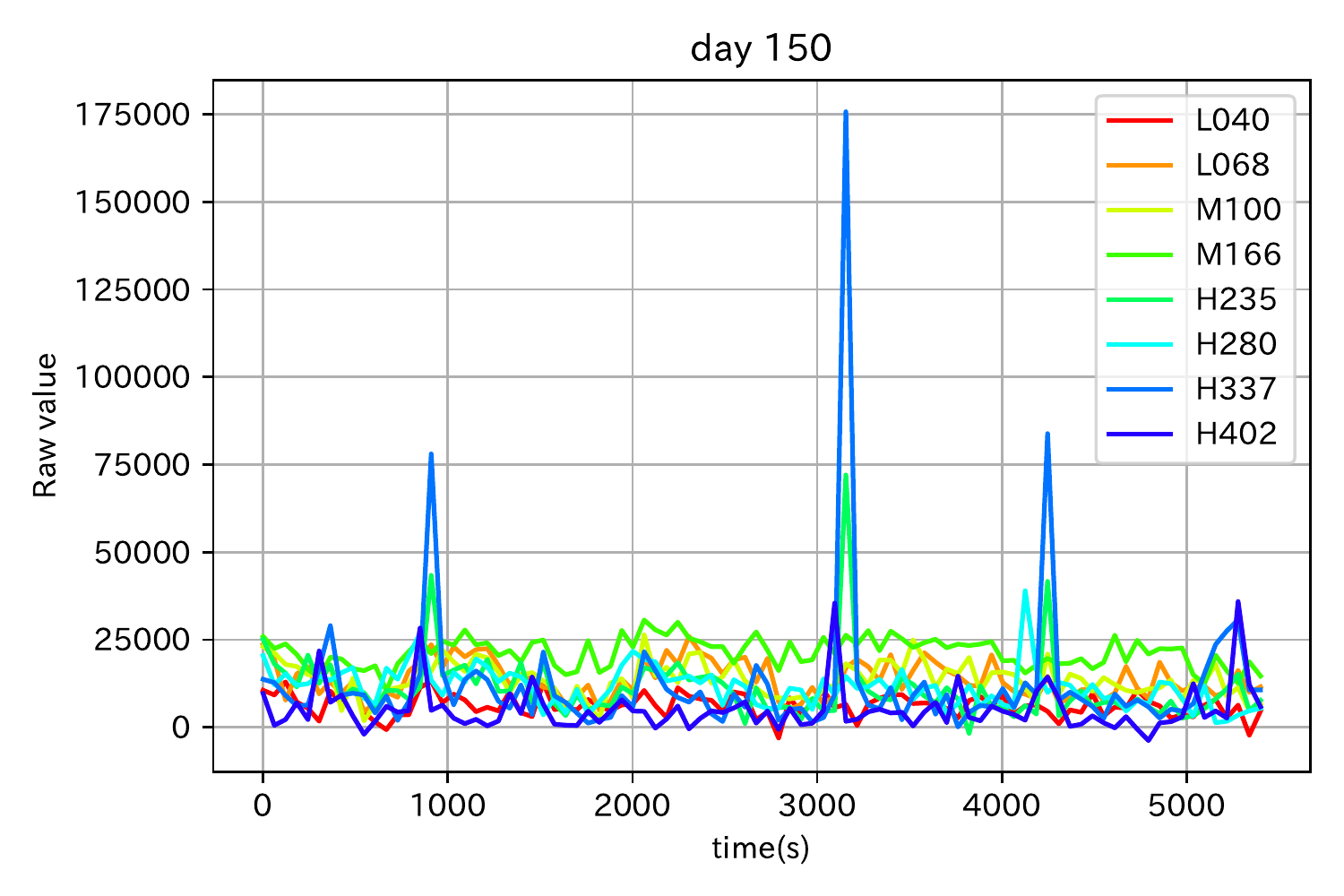}
  \end{minipage}
 \end{tabular}
 \vspace*{-3mm}
 \caption{(left) Simulated TOD for 40~GHz in LFT separately for each component and their
 total. (right) Simulated TOD in day 150 for some representative frequency (L/M/H is
 for LFT/MFT/HFT), and the numbers are for the frequency in GHz.}
\label{f3-1-4}
\end{figure}

We started the mock observation from the vernal equinox for one year for all frequency
bands. We focus on days 150 and 270 with the largest variation, in which the scan
includes a pole of the CMB dipole and the Galactic center, respectively. In
Fig.~\ref{f3-1-4}, the left panel shows a 6~s TOD in the 40~GHz band of LFT separately
for each component. The total power is governed by the white noise centered at zero, the CR
glitches positively biased but not time-resolved, and the T-to-P leakage modulated by
2$f_{\mathrm{HWP}}$. The right panel shows a 6000~s TOD in some representative
bands. Passages of the Galactic plane are apparent particularly in high frequency bands.

We calculated the information entropy for the simulated TOD in each day and each band to
be 13--16~bits. Here, the information entropy is a quantity to describe how random the data are.  If a
variable $X$ takes a value {$x_{1}$, ..., $x_{n}$} with a probability {$P (x_{1})$, ...,
$P (x_{n})$}, the information entropy of $X$ is defined as
\begin{equation}
 I(X)=-\sum_{i=1}^{n}P(x_i)\log{P(x_i)}.
\end{equation}
We calculated the probability $ P(x_i)$ based on the normalized histogram of simulated
TOD. This is losslessly compressible as the TOD of CMB observations have temporal correlations,
which we should exploit. Fortunately, some major components contributing to the
information entropy are well-behaved and can be fitted with a simple model.  We downlink
both the best-fit parameters and the residual TOD so that we can fully reproduce the
observed TOD on the ground.

We first tried the linear polynomial fitting employed in many compression
algorithms for general use. We optimized the fitting parameters, but this did not work
very well. This is because the major component of TOD, the T-to-P leakage, is a cosine
curve in nature and is not appropriate to fit to a low-order polynomial. We next tried
differential compression, in which the differential of the adjacent two samples are
downlinked along with the first sample in a block, so that all samples in the block can
be reproduced on the ground. This is effective in removing slowly changing components,
and yielded sufficient compression. We also exploit the primarily cosine nature of the TOD
and fitted it with a function
\begin{math}
 A\cos{(2\omega_{\mathrm{HWP}}t+\phi)}+c,
\end{math}
in which $A$, $\phi$ and $c$ are free parameters. The cosine term is intended for the
T-to-P leakage component, and the constant term is for slowly varying foreground components
and the bias of the CR glitches. We fit a certain length of TOD and subtract the model
from the observed TOD. The optimum fitting length (1000 samples $=$ 6.5~s) was chosen by
the trade-off between long to constrain the T-to-P leakage parameters and short to catch up
with the rapid changes of the foregrounds through Galactic plane passages and the change
of $\phi$ as the satellite scans.

\begin{figure}[hbtp]
  \begin{center}
  \includegraphics[keepaspectratio,width=\columnwidth]{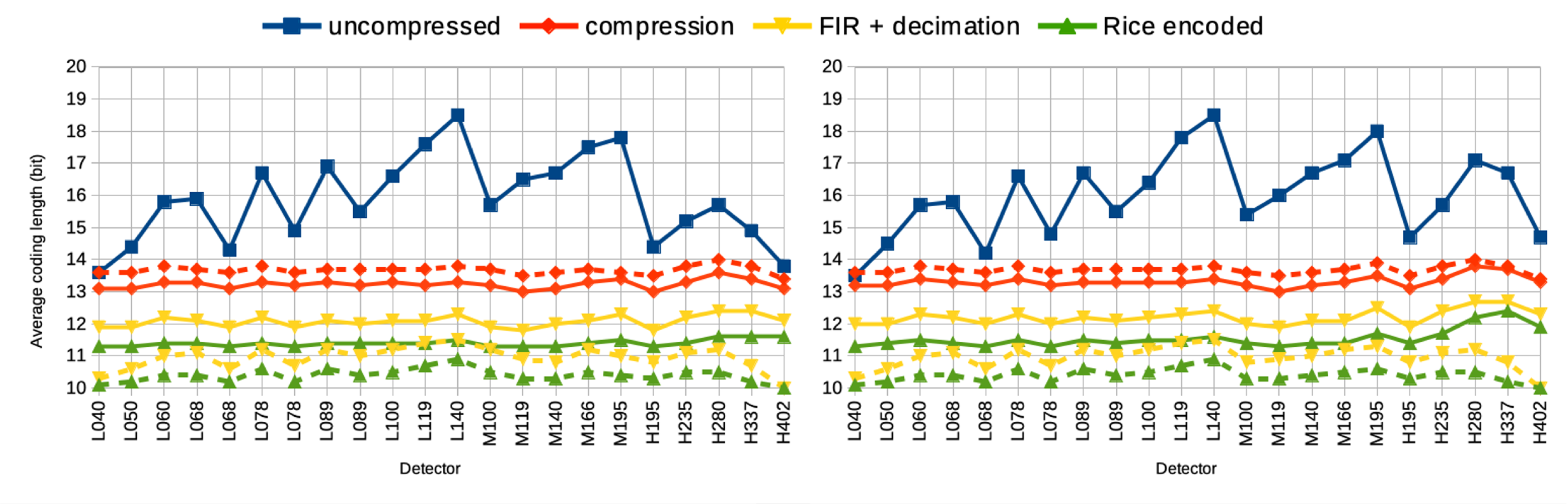}
  \end{center}
\caption{Average encoded length for day 150 (left) and 270 (right) for each frequency. Solid lines represent the results for fitting subtraction and dashed lines for differentiation.}
\label{f3-2}
\end{figure}

We actually applied all these algorithms (polynomial fitting, differentiation, and
cosine fitting) to the simulated TOD. We then calculated the residual TOD after fitting
or differentiation, which was filtered with the FIR and decimated to 19~Hz. We encoded
the 19~Hz residual TOD using a lossless compression (Rice encoding \cite{Rice1971}) and
added the bit length of the fitting parameters to derive the average bit length. The
result for the cosine fitting is shown in Fig.~\ref{f3-2} for day~150 and 270. Without
fitting, the 153~Hz TOD (``uncompressed'') can be encoded only to an average length of
19~bits. After subtracting the best-fit model (solid line) or calculating differentiation (dashed line), the average
length decreases to 13.5~bits at the maximum (``compression'').
After FIR filtering and decimation to
19~Hz (``FIR $+$ decimation'') and Rice encoding (``Rice encoded''), the average length
is 12~bits, achieving the compression ratio of 50\%. Two high-frequency channels using fitting algorithm exceed
the 12~bit limit in day~270 . However, the length averaged over the entire frequency band
weighted by the number of channels is below 12~bit, which is sufficient for our
needs. The differentiation algorithm also yielded a similar compression rate ($\sim$
50\%) with a block size of 24 samples. Each has its own advantages. The trade-off
between them with further tuning under onboard resource constraints is a work to be
done.

\section{Conclusion}\label{s4}
In this study, we calculated both the requirement and the achievable value of the data
compression ratio in the orbit. In the case of lossless compression, the requirement was
derived to be 0.50 from the data rate per channel, the number of channels, the bit
length, and the exposure. The achievable value was estimated from the simulated TOD to
be as realistic as possible using the currently available design values of the
experiment. We investigated several representative algorithms and demonstrated that
those based on the differentiation or the cosine fitting of the TOD meet the requirement.

\begin{acknowledgements}

Some of the results in this paper have been derived using the \texttt{healpy} and
\texttt{HEALPix} package. This work was supported by JSPS KAKENHI Grant Number
JP18K03715. We thank Francesco Piacentini, Tijmen de Haan and Tomotake Matsumura for improving the manuscript.
\end{acknowledgements}


\begin{thebibliography}{}
 \bibliographystyle{spphys}
 \bibitem{Adam2016b} R. Adam et al., A\&A, 594, A8 (2016)
 \bibitem{Planck2018} N. Aghanim et al., A \& A, 641, A3 (2020)
 \bibitem{Dobbs2012} M. Dobbs et al., Review of Scientific Instrument, 83, 073113 (2012)
 \bibitem{Gaztanaga2001} E. Gaztanaga et al., MNRAS, 302, 1 (2001)
 \bibitem{Ghigna2020} T. Ghigna, Ph.D. thesis, University of Oxford (2020)
 \bibitem{Hazumi2020} M. Hazumi et al., Proc of SPIE, 11443, 114432F (2020)
 \bibitem{Maris2000a} M. Maris et al., A\&A, 147, 1 (2000)
 \bibitem{Maris2004} M. Maris et al., A\&A, 414, 2 (2004)
 \bibitem{Montgomery2020} J. Montgomery, Ph.D. thesis, McGill University (2021)
 \bibitem{Montier2020} L. Montier et al., Proc of SPIE, 11443, 114432G (2020)
 \bibitem{Rice1971} R. Rice and J. Plaunt, IEEE Trans. on Comm. Tech., 19, 6 (1971)
 \bibitem{Sekimoto2020} Y. Sekimoto et al., Proc of SPIE, 11453, 1145310 (2020)
 \bibitem{Shannon1948} C. E. Shannon, Bell System Technical Journal, 27, 623 (1948)
 \bibitem{Stever2021} S. L. Stever et al., JCAP, 09, 013 (2021)
 \bibitem{Thorne2016} B. Thorne et al., MNRAS, 469, 3, 2821 (2016)
 \bibitem{Tominaga2020} M. Tominaga et al., Proc of SPIE, 11453, 114532H (2020)
\end{thebibliography}

\end{document}